\documentclass{article}
\usepackage{graphicx} % Required for inserting images
\usepackage{authblk}
\usepackage{amsmath}
\usepackage{multirow}
\usepackage[sort, comma, numbers]{natbib}
\usepackage{pdflscape}
\usepackage[skip=5pt plus1pt, indent=40pt]{parskip}

\usepackage{amssymb,amsfonts}
\usepackage{textcomp}
\usepackage{xcolor}

\usepackage{geometry}
\usepackage{fancyhdr}
\title{Literary and Colloquial Dialect Identification for Tamil using Acoustic Features}
\author[1]{M. Nanmalar}
\author[1]{P. Vijayalakshmi}
\author[1]{T. Nagarajan}
\affil[1]{Sri Sivasubramaniya Nadar College of Engineering, Chennai}
\date{}

\usepackage{titling}
\begin{document}
\maketitle

\begin{center}
\section*{Abstract}
\end{center}
The evolution and diversity of a language is evident from it's various dialects. If the various dialects are not addressed in technological advancements like automatic speech recognition and speech synthesis, there is a chance that these dialects may disappear. Speech technology plays a role in preserving various dialects of a language from going extinct. In order to build a full fledged automatic speech recognition system that addresses various dialects, an Automatic Dialect Identification (ADI) system acting as the front end is required. This is similar to how language identification systems act as front ends to automatic speech recognition systems that handle multiple languages. The current work proposes a way to identify two popular and broadly classified Tamil dialects, namely literary and colloquial Tamil. Acoustical characteristics rather than phonetics and phonotactics are used, alleviating the requirement of language-dependant linguistic tools. Hence one major advantage of the proposed method is that it does not require an annotated corpus, hence it can be easily adapted to other languages. Gaussian Mixture Models (GMM) using Mel Frequency Cepstral Coefficient (MFCC) features are used to perform the classification task. The experiments yielded an error rate of 12\%. Vowel nasalization, as being the reason for this good performance, is discussed. The number of mixture models for the GMM is varied and the performance is analysed.

\vspace{0.25cm}
\begin{center}
\textbf{Keywords}: \textit{dialect identification, corpus, GMM, MFCC}
\end{center}

\section{Introduction}\label{intro}
Automatic Language Identification (LID) is the problem of identifying the language from a small segment of speech uttered by an unknown speaker, with or without any prior knowledge about the target languages. Automatic LID systems can be used as a front end for Automatic Speech Recognition (ASR) and Automatic Machine Translation (AMT) systems. For example, in public service points like airports and railway stations, non native speakers buying tickets use the enquiry systems using these ASR systems. An Automatic Dialect Identification (ADI) system is similar to an LID in that, it deals with the dialects of a single language instead of multiple languages. Here, our intention is to classify Literary Tamil (LT) and Colloquial Tamil (CT). Literary Tamil is an ancient form of Tamil that is used in spoken as well as written forms. Colloquial Tamil is the modern spoken form of Tamil that is not usually used in written form. Though they belong to the same language, there is a considerable difference between LT and CT, and are hence different dialects of the Tamil language.

The problem statement of the current work is, given a speech utterance, without any prior phonotactic knowledge about both LT and CT, the system is to classify whether the test utterance belongs to LT or CT. The current work would act as one of the main components of a full-fledged dialect identification system. It can also be used in a dialect conversion system. The system will be helpful for the common man who is not accustomed to literary Tamil, but is very familiar with the colloquial one. The task is not trivial because it involves dialects of the same language. The similarity complicates the classification problem. In the current work, we propose that Gaussian Mixture Models (GMM) can aptly model the acoustic characteristics of each of these dialects and classify them. We detail how GMM is helpful in building a dialect classifier with a small error rate and how vowel nasalization, as an important cue for classification, has aided in achieving the performance.

In \cite{torres2002approaches}, GMM Tokenization along with Shifted Delta Cepstra (SDC) feature vectors, which provide additional temporal information about the speech, is used. This reduced the error rate of classifying Arabic and Vietnamese languages when compared to using MFCC features. In \cite{manchala2014gmm}, Mel Frequency Cepstral Coefficients (MFCC) along with formant frequencies extracted using LP spectrum are combined to form a new feature and a GMM based classifier to classify 10 languages in the OGI corpus is built. The performance of the system is compared with that of a system that uses just MFCC features, and the authors claim better performance when the new feature is used. In \cite{zissman1996comparison}, GMM and PRLM techniques for LID are compared. Both GMM and HMM models are compared in \cite{zissman1993automatic} and it is found that the results are comparable. The authors also state that, the Lincoln algorithm stated in their work requires no hand labeled training data, as compared to systems that require them. This would mean that the work can be easily expanded to other languages. The authors suggest that future work can focus on building over simple statistical approaches, as compared to sophisticated methods where phonological knowledge is required. In \cite{chen2001automatic}, a multi-accent mandarin corpus developed for four different accents is used to build GMMs which classify these accents. Hence, we find that GMM based methods that use additional features with MFCC as in \cite{torres2002approaches} \cite{manchala2014gmm}, or just MFCC features, can be used to obtain good classification accuracy.

The organisation of the paper is as follows: The similarities and differences of literary Tamil and colloquial Tamil are detailed in Section \ref{linguistic_nature}. Corpus details are provided in section \ref{corpus}. Section \ref{nasalization} describes the challenges in identifying dialects and how it can be resolved by finding distinctive cues such as vowel nasalization. Section \ref{system} details the GMM training process and an experiment performed to identify vowel nasalization. Section \ref{results} discusses the results obtained.

\section{Linguistic Nature Of Tamil}\label{linguistic_nature}

As of 2017, 70 million people read, write or speak Tamil. All of these people come under the use of different forms of Tamil, called dialects. These dialects stem from culture, region and religion. Tamil dialects can be broadly classified into literary and colloquial Tamil. The differences between literary and colloquial forms of Tamil are as mentioned below:
\begin{itemize}
    \item While literary Tamil has a standard orthographic representation (written form) and spelling, colloquial Tamil does not possess an official standard \cite{keane2004tamil}
    \item The literary form possess standard grammar, while the colloquial form does not
\end{itemize}

Within this scenario, if comparisons have to be made between literary and colloquial Tamil, they can be made in three ways:
\begin{itemize}
    \item Word level changes: The whole word may be different
    \item Phoneme level changes: Some phonemes of the word may be different
    \item Vowel nasalisation: Vowels may be nasalized
\end{itemize}

\begin{figure}[htbp]
\centerline{\includegraphics[width=0.60\textwidth]{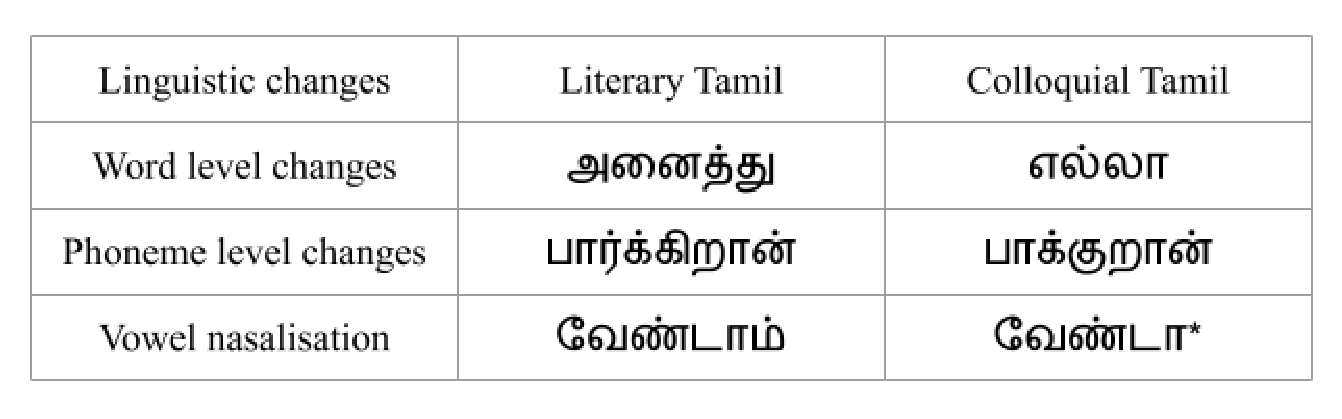}}
\caption{Examples showing differences between literary and colloquial Tamil}
\label{fig_examples}
\end{figure}

This comparison is explained figuratively in Fig.\ref{fig_examples}. Hence, we see that methods that use orthographic representations or knowledge thereof, may not produce satisfying results. In view of this, a GMM based method that considers only the acoustic characteristics of the signal can be used to obtain better results.

\section{Corpus}\label{corpus}

\subsection{Existing Corpus}
Existing work in LID literature use the Oregon Graduate Institute Multi-Language Telephone Speech (OGI-TS) corpus as mentioned in \cite{manchala2014gmm} \cite{torres2002approaches}. The OGI corpus contains more than ten languages and their dialects. But the corpus contains telephonic speech. This makes it unsuitable for training models that will be used in all environments. Also, this corpus has a limited vocabulary. Hence, it was decided to create a corpus that is customized to deal with literary and colloquial dialects.

\subsection{Corpus created in the current work}
The corpus that is created in the current work has three main features,
\begin{itemize}
    \item they are collected from multiple speakers
    \item they are collected in multiple noise conditions - quiet lab environment, lab environment with ac on, lab environment with the fan on.
    \item they are also collected using different microphones namely, a condenser and a dynamic microphone, simultaneously.
\end{itemize}

\subsection{Text Corpus}
The text corpus contains text for both literary and colloquial Tamil. The literary Tamil text sentences are collected from short stories, literatures and proverbs. The sources that provide literary text data are abundantly available on the world wide web. The colloquial text sources are not as abundant as the literary ones. Hence, the authors had to rely on novels, magazine articles and interviews that appear on magazines. It is not assured that the quality of these sources will be satisfactory, hence a level of verification and correction is required. The type of corrections maybe removal of literary text that may be present in between, or just phonetic or spelling corrections to bring to an approximately standard form.

\subsection{Speech Corpus}
The text collected in the previous step is used as a source to develop the speech corpus. The speech for the speech corpus is contributed by people of different age groups, namely students and faculties in the institution. Both male and female speech are collected. Almost all the speakers contributed both literary and colloquial speech. The collection of colloquial speech data is challenging as there is no standardized form of writing and interpretation of colloquial Tamil text. The speech corpus is recorded sentence by sentence. The average length of each of these sentences is 5 seconds. It must be stated that this is not a parallel corpus of literary and colloquial Tamil. The text sentences are independently collected. Details of the speech corpus collected are mentioned in Table \ref{tab_corpus}.

\begin{table}[htbp]
\caption{Details of the corpus}
\vspace{0.25cm}
\begin{center}
\begin{tabular}{|l|l|l|ll}
\cline{1-3}
\multicolumn{1}{|c|}{\textbf{Corpus detail}}      & \multicolumn{1}{c|}{\textbf{LT}} & \multicolumn{1}{c|}{\textbf{CT}} &  &  \\ \cline{1-3}
Total duration of speech corpus in hours & 8.03                               & 5.48                               &  &  \\ \cline{1-3}
Total number of speakers                 & 79                               & 77                               &  &  \\ \cline{1-3}
Total number of male speakers            & 24                               & 25                               &  &  \\ \cline{1-3}
Total number of female speakers          & 55                               & 52                               &  &  \\ \cline{1-3}
\end{tabular}
\label{tab_corpus}
\end{center}
\end{table}

There is no overlap between the train and test speakers. The details of how the data is split between train and test is detailed in Table \ref{tab_traintest}. The condenser microphone used in the setup is Rode-NT1A and the dynamic microphone used in the setup is Ahuja 100XLR.

\begin{table}[htbp]
\caption{Train and Test split up}
\vspace{0.25cm}
\begin{center}
\begin{tabular}{|l|l|l|l|}
\hline
\textbf{Detail}                                  & \textbf{Train/Test} & \multicolumn{1}{c|}{\textbf{LT}} & \multicolumn{1}{c|}{\textbf{CT}} \\ \hline
Total duration in hours          & Train      & 6'19"                            & 3'55"                            \\ \cline{2-4} 
                                                 & Test       & 1'43"                            & 1'52"                            \\ \hline
Total number of male speakers   & Train      & 21                               & 20                               \\ \cline{2-4} 
                                                 & Test       & 3                                & 6                                \\ \hline
Total number of female speakers & Train      & 43                               & 37                               \\ \cline{2-4} 
                                                 & Test       & 12                               & 15                               \\ \hline
\end{tabular}
\label{tab_traintest}
\end{center}
\end{table}

\section{Nasalization Analysis} \label{nasalization}
\subsection{Challenges in dialect classification}
In text independent LID tasks, phonemes and other sub-word units alone may not be sufficient enough to classify between languages. Hence it is better to look at whole sentences and derive acoustic signatures from them to identify a language. To obtain acoustic signatures, the following parameters will be required \cite{muthusamy1994reviewing}:
\begin{itemize}
    \item Phonetics - frequency of occurence of phonemes
    \item Prosodics - duration and intonation of the phonemes
    \item Phonotactics - grammar rules for the phonemes
\end{itemize}

The above characteristics can be used to identify dialects as well. However, in case of dialect identification the task is more complicated since there are a lot of similarities in the use of phonetics across dialects. The words with same meaning used in two different dialects often do not change. In the case of phonetics also, not much change can be guaranteed. Besides these phonetical changes are not standardized as well. Whether this can be used as a distinctive cue is doubtful. Hence we propose that vowel nasalization can be a distinctive cue to find the difference between LT and CT. 

\subsection{Vowel Nasalization}
There are three kinds of vowels that are produced when speaking in Tamil. These are, 
\begin{itemize}
    \item Oral vowel: Resonance occurs only in the oral cavity
    \item Nasal vowel: Velum lowers and hence resonance occurs in both the oral and nasal cavity simultaneously
    \item Nasalized vowel: Vowel that is nasalized due to coarticulation induced by the neighboring phoneme. In Tamil, this is produced due to assimilation of nasal consonants. It causes an increase in vowel height.
\end{itemize}

Nasalization can occur in various degrees producing, lightly or highly nasalized vowels. If a vowel and nasal consonant occur together, it automatically causes the vowel to be nasalized \cite{keane2004tamil}. If the vowel is short, then the quality of nasalization varies. Such nasalization also causes the nasal consonant to be eliminated in the colloquial form. This stems from the empirical observation of literary and colloquial Tamil. 

\section{System Building} \label{system}
In the current work only acoustic models are used and no langugage models are used. Mel Frequency Cepstral Coefficient (MFCC) features, that contain 13 each of static, velocity and acceleration coefficients (39 dimensions in total), is used to model a GMM. A GMM is a probabilistic model where weighted component gaussian densities are combined. Eq.\ref{eq1} describes GMM. Here, M represents mixture weights, $\textbf{x}$ represents data vector from 39 dimension and $g(\textbf{x}|\mu_i,\sigma_i)$ represents the component Guassian densities. Each density component contains its own mean, covariance and mixture weight. MFCC is a perceptual filtering technique which is used commonly for extraction of features from speech signals.

\begin{equation}
p(\textbf{x}|\lambda) = \sum_{i=1}^{M} w_i g(\textbf{x}|\mu_i,\sigma_i) \label{eq1}
\end{equation}

A GMM is built for LT and CT separately using the 39 dimensional MFCC features mentioned above. Cepstral mean subtraction is performed on the features in order to reduce the influence caused by the environment in which the audio is recorded for training. The same is applied during the testing phase as well. The number of mixture components of the GMM is varied and the performance is analysed. Depending on the data, the number of GMM components/mixtures can affect the performance of the system \cite{chen2001automatic}. But as the number of mixture components are increased, the computational complexity also increases. The best performance of the system is 87\% and it can be justified using the vowel nasalisation experiment explained in \ref{vowel_nasalisation}. The details of the performance are discussed in section \ref{results} 

\subsection{Experiment on vowel nasalisation}\label{vowel_nasalisation}
As mentioned in the previous section, a vowel takes the form of a nasalized vowel because of the assimilation of the nasal consonant right next to it. A point to note is that the degree of nasalisation differs between literary Tamil and colloquial Tamil. In the case of literary Tamil, it is lightly nasalized, and in the case of colloquial Tamil it is highly nasalized.
\begin{itemize}
    \item In literary Tamil, when the vowel is nasalized by the nasal consonant that follows it, the influence of both the vowel and the consonant sounds is present
    \item In colloquial Tamil, when the vowel is nasalized by the nasal consonant that follows it, it assimilates into a nasalized vowel, eliminating the nasal consonant in the process
\end{itemize}

Hence the degree of nasalization is high in colloquial Tamil, due to the elimination of nasal consonant. This concept is proved by a simple experiment where the characteristics of a nasal vowel is analysed in both literary Tamil and colloquial Tamil, as explained in detail below.

Mostly vowel nasalisation happens in the word boundary. In Tamil, nasal consonants that occur in word boundary are the phonemes 'm' and 'n'. In this experiment, we acoustically analyse the long vowel 'aa' in context with 'm' and 'n'. We analyse its degree of nasalization and its correspondence with literary Tamil and colloquial Tamil. We record single words which contain 'm' or 'n' as the end phoneme, and preceeded by 'aa', to show the nature of nasalisation in a single vowel. The segment at the end of the word is taken for analysis. The vowel nasal segment is divided into 20 ms frames, and the formants are extracted from the LP spectrum.

\begin{figure}[htbp]
\centerline{\includegraphics[width=0.85\textwidth]{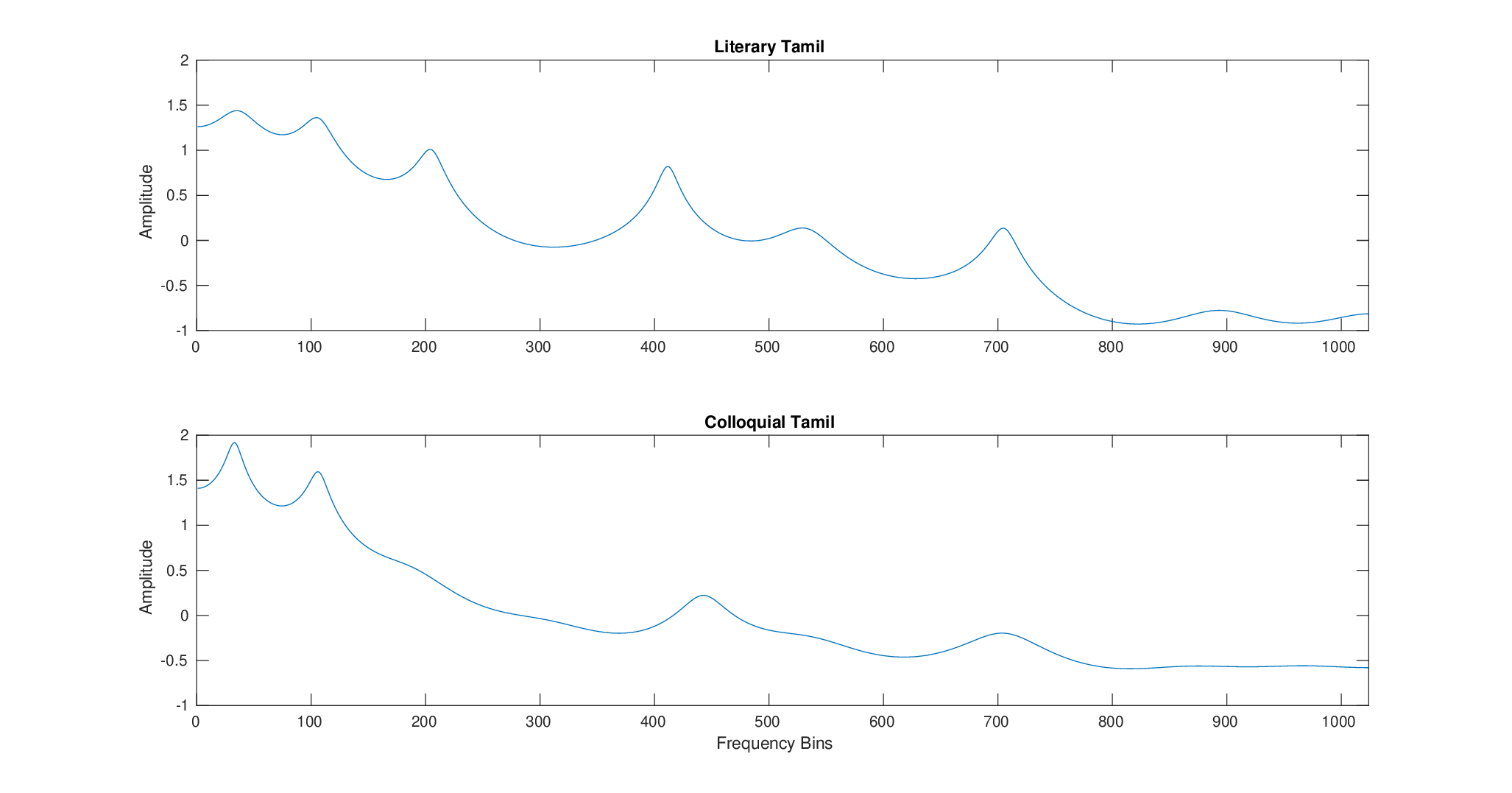}}
\caption{Comparison of the degree of nasalization of vowel 'aa' in literary and colloquial Tamil}
\label{fig_nasalized}
\end{figure}

In \cite{vijayalakshmi2007acoustic}, it is shown that when the vowel 'aa' followed by one of the two nasal consonants 'm' or 'n' exhibits nasalization, with a formant peak around 250Hz. In our experiment, we observed the same characteristic. In order to check the degree of nasalization, the magnitude of the formant peak is measured. When empirically measuring the magnitude of the peaks of nasalized vowels in colloquial and literary Tamil, we observe that colloquial Tamil consistently has stronger formants compared to literary Tamil. This phenomenon is graphically represented in Fig. \ref{fig_nasalized}

\section{Results and Discussions} \label{results}

The details of the test corpus is listed in Table \ref{tab_traintest}. Each test utterance contains a sentence of speech. Testing is performed sentence by sentence. The maximum likelihood of a particular test utterance is calculated with respect to GMMs built for literary and colloquial Tamil. The one which has higher likelihood is chosen as the class for a particular test utterance. It must be noted that the average length of a test utterance is 5s. Even with such short utterences, the system is able to classify the literary and colloquial forms. Even though there are a lot of phonetic similarities between literary and colloquial forms, the results are promising.

The mixture components of the GMM are varied and the performance is analysed in each case. It is observed that for the corpus as detailed in Table \ref{tab_corpus} and \ref{tab_traintest}, best classification performance of 88\% is achieved with 256 mixture components. In this case, considering literary Tamil as the positive condition, the precision, recall and F1 measure are computed and found to be 0.82, 0.89 and 0.85 respectively. The results are detailed in Table \ref{tab_resultsmix}

\begin{table}[htbp]
\caption{Performance Comparison for Various Mixture Components}
\vspace{0.25cm}
\begin{center}
\begin{tabular}{|c|c|ll}
\cline{1-2}
\textbf{Number of Components} & \textbf{Performance (\%)} &  &  \\ \cline{1-2}
16                            & 84.89                     &  &  \\ \cline{1-2}
32                            & 87.06                     &  &  \\ \cline{1-2}
64                            & 86.40                     &  &  \\ \cline{1-2}
128                           & 87.72                     &  &  \\ \cline{1-2}
256                           & 88.00                     &  &  \\ \cline{1-2}
\end{tabular}
\label{tab_resultsmix}
\end{center}
\end{table}

From Table \ref{tab_resultsmix}, it is seen that, as the number of mixtures or, component gaussians increase, the performance also increases. But as the number of component gaussians are increased, the computational complexity also increases. For example, on a computer runnning on i5 processor with 8GB ram, it took 4 days for the models to be built. Since the performance has not increased considerably, we may trade off for better computational complexity, when there is a need.

\section{Conclusion} \label{conclusion}
A GMM based classifier for idenfiying literary and colloquial Tamil is built. GMMs have been used in the past for language and dialect identification systems, but it has not been used in classifying Tamil dialects. Compared to other dialect classification attempts as in \cite{chen2001automatic} and \cite{zissman1993automatic}, the proposed method performs better with an error rate of just 12\%. Considering the similarities between literary and colloquial Tamil, this is a commendable score. The justification of this performance is provided by the vowel nasalization analysis.

The biggest advantage of the proposed method is that no transcription is needed. This means that, the methodology can be extended to other dialects in Tamil as well. The process is efficient in that, it eliminates the need for time-aligned transcription of the wave files, and language models. The dataset can be further extended by simply requesting users/participants to speak in a given dialect. This will increase the taining data, and cause an increase in performance. The solution achieved is simple and efficient.

In the future, dialect identification systems can be built with transcriptions and the results can be compared with the current work. Deeper analysis on the importance of nasalization in dialect identification can be performed.

% \section*{References}
%\bibliographystyle{ieeetr}
%\bibliography{myref}

\end{document}